\documentclass{fac}
\usepackage{graphicx}
\usepackage{amssymb}
\usepackage{amsfonts}
\usepackage{amsmath}
\usepackage{dsfont}
\usepackage{MnSymbol}
\usepackage{mathtools}
\usepackage{tikz}
\usetikzlibrary{calc,decorations.pathmorphing,shapes}
\ifprodtf \else \usepackage{latexsym}\fi

\newcommand\black{\ensuremath{\blacktriangleright}}
\newcommand\white{\ensuremath{\vartriangleright}}

\newif\ifamsfontsloaded
\ifprodtf
  \newcommand\whbl{\white\kern-.1em--\kern-.1em\black}
  \newcommand\blwh{\black\kern-.1em--\kern-.1em\white}
  \newcommand\blbl{\black\kern-.1em--\kern-.1em\black}
  \newcommand\whwh{\white\kern-.1em--\kern-.1em\white}
  \amsfontsloadedtrue
\else
  \checkfont{msam10}
  \iffontfound
    \IfFileExists{amssymb.sty}
      {\usepackage{amssymb}\amsfontsloadedtrue
       \newcommand\whbl{\white\kern-.125em--\kern-.125em\black}%
       \newcommand\blwh{\black\kern-.125em--\kern-.125em\white}%
       \newcommand\blbl{\black\kern-.125em--\kern-.125em\black}%
       \newcommand\whwh{\white\kern-.125em--\kern-.125em\white}}
      {}
  \fi
\fi

\title[Draft of Weakly True Concurrency and Its Logic]
      {Weakly True Concurrency and Its Logic}

\author[Yong Wang]
    {Yong Wang\\
     College of Computer Science and Technology,\\
     Beijing University of Technology, Beijing, China\\
     }

\correspond{Yong Wang, Pingleyuan 100, Chaoyang District, Beijing, China.
            e-mail: wangy@bjut.edu.cn}

\pubyear{2016}
\pagerange{\pageref{firstpage}--\pageref{lastpage}}

\begin{document}
\label{firstpage}

\makecorrespond

\maketitle

\begin{abstract}
We firstly find the existence of silent event $\tau$ in true concurrency (named weakly true concurrency) by defining weakly true concurrent behaviors and weakly true concurrent logics. Based on Paolo Baldan and Silvia Crafa's comprehensive work on (strongly) true concurrency, we find the correspondence between weakly true concurrent equivalences and weakly true concurrent logics.
\end{abstract}

\begin{keywords}
True Concurrency; Modal Logic; Behaviorial Equivalence; Prime Event Structure; Silent Event; Weakly True Concurrency
\end{keywords}

\section{Introduction}{\label{intro}}

In recent years, there have emerged several logics on true concurrency, including EIL (Event Identifier Logic) \cite{RL1} \cite{RL2}, SFL (Separation Fixpoint Logic) and TFL (Trace Fixpoint Logic) \cite{SFL}, and Paolo Baldan and Silvia Crafa's comprehensive work \cite{LTC1} \cite{LTC2} on (strongly) true concurrency. We will not enumerate all work on true concurrency, but, all the work neglects the silent event $\tau$ in true concurrency background, which $\tau$ is called silent step in (interleaving) bisimilarity \cite{ALNC} \cite{CC}.

\begin{figure}
  \centering
  \includegraphics{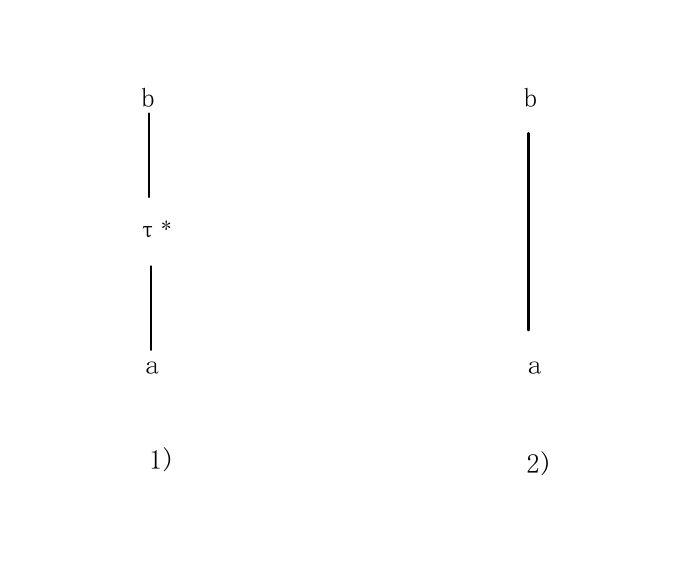}
  \caption{}
  \label{example}
\end{figure}

We consider the two prime event structures (PESs) Fig.\ref{example}.1) and Fig.\ref{example}.2), which are denoted as the CCS processes $a.\tau^*.b$ and $a.b$. There exists several silent events $\tau^*$ in Fig.\ref{example}.1), which is invisible from the outside world. Since any strongly behaviorial equivalences (such as interleaving bisimilarity, pomset bisimilarity, step bisimilarity, history-preserving bisimilarity and hereditary history-preserving bisimilarity) do not distinguish internal invisible and external visible events, the two PESs in Fig.\ref{example} are not equivalent modulo any (strongly) concurrent bisimilarity. But, if we consider silent event $\tau$ in concurrent bisimilarity, just as (interleaving) bisimilarity done, which is called weak (interleaving) bisimilarity, we will establish the concept of weakly true concurrent bisimilarity (including weak pomset bisimilarity, weak step bisimilarity, weak history-preserving bisimilarity and weak hereditary history-preserving bisimilarity). In fact, the two PESs in Fig.\ref{example} are equivalent modulo any weakly concurrent bisimilarity.

We introduce silent event $\tau$ into true concurrency based on Paolo Baldan and Silvia Crafa's comprehensive work \cite{LTC1} \cite{LTC2} on (strongly) true concurrency, just because \cite{LTC1} \cite{LTC2} unified several (strongly) concurrent bisimilarity under one framework of modal logic, and it is natural to extend it to weakly concurrent bisimilarity under one modal logic, and for this, we believe. Although the extension looks like somewhat a trivial work, we just process carefully to make silent event $\tau$ really keep silent.

The rest sections are organized as follows: in section \ref{wtc}, we extend PES with silent event $\tau$ and introduce the concepts of several weakly concurrent bisimilarity; in section \ref{logic}, we extend $\mathcal{L}$ to $\mathcal{L}^{\tau}$, and prove that $\mathcal{L}^{\tau}$ induces weak hereditary history-preserving bisimilarity; we extend fragments in $\mathcal{L}$ to those in $\mathcal{L}^{\tau}$ in section \ref{frag}; and also, we consider recursion in section \ref{recur}; finally, we conclude this paper in section \ref{conc}.

\section{Weakly True Concurrency}{\label{wtc}}

In this section, we extend prime event structure with silent event $\tau$, and explain the concept of weakly true concurrency, i.e., concurrent behaviorial equivalence with considering silent event $\tau$.

\subsection{Event Structure with Silent Event $\tau$}

We give the definition of prime event structure (PES) \cite{ES1} \cite{ES2} \cite{CM} extended with the silent event $\tau$ as follows.

\textbf{Definition \ref{wtc}.1 (Prime Event Structure with Silent Event).} Let $\Lambda$ be a fixed set of labels, ranged over $a,b,c,\cdots$ and $\tau$. A ($\Lambda$-labelled) prime event structure with silent event $\tau$ is a tuple $\mathcal{E}=\langle E, \leq, \sharp, \lambda\rangle$, where $E$ is a denumerable set of events, including the silent event $\tau$. Let $\hat{E}=E\backslash\{\tau\}$, exactly excluding $\tau$, it is obvious that $\hat{\tau^*}=\epsilon$, where $\epsilon$ is the empty event. Let $\lambda:E\rightarrow\Lambda$ be a labelling function and let $\lambda(\tau)=\tau$. And $\leq$, $\sharp$ are binary relations on $E$, called causality and conflict respectively, such that:

\begin{enumerate}
  \item $\leq$ is a partial order and $\ulcorner e \urcorner = \{e'\in E|e'\leq e\}$ is finite for all $e\in E$. It is easy to see that $e\leq\tau^*\leq e'=e\leq\tau\leq\cdots\leq\tau\leq e'$, then $e\leq e'$.
  \item $\sharp$ is irreflexive, symmetric and hereditary with respect to $\leq$, that is, for all $e,e',e''\in E$, if $e\sharp e'\leq e''$, then $e\sharp e''$.
\end{enumerate}

Then, the concepts of consistency and concurrency can be drawn from the above definition:

\begin{enumerate}
  \item $e,e'\in E$ are consistent, denoted as $e\frown e'$, if $\neg(e\sharp e')$. A subset $X\subseteq E$ is called consistent, if $e\frown e'$ for all $e,e'\in X$.
  \item $e,e'\in E$ are concurrent, denoted as $e\parallel e'$, if $\neg(e\leq e')$, $\neg(e'\leq e)$, and $\neg(e\sharp e')$.
\end{enumerate}

\textbf{Definition \ref{wtc}.2 (Configuration).} Let $\mathcal{E}$ be a PES. A (finite) configuration in $\mathcal{E}$ is a (finite) consistent subset of events $C\subseteq \mathcal{E}$, closed with respect to causality (i.e. $\ulcorner C\urcorner=C$). The set of finite configurations of $\mathcal{E}$ is denoted by $\mathcal{C}(\mathcal{E})$. We let $\hat{C}=C\backslash\{\tau\}$.

A consistent subset of $X\subseteq E$ of events can be seen as a pomset. Given $X, Y\subseteq E$, $\hat{X}\sim \hat{Y}$ if $\hat{X}$ and $\hat{Y}$ are isomorphic as pomsets. In the following of the paper, we say $C_1\sim C_2$, we mean $\hat{C_1}\sim\hat{C_2}$.

\textbf{Definition \ref{wtc}.3 (Weak Pomset Transitions and Weak Step).} Let $\mathcal{E}$ be a PES and let $C\in\mathcal{C}(\mathcal{E})$, and $\emptyset\neq X\subseteq \hat{E}$, if $C\cap X=\emptyset$ and $\hat{C'}=\hat{C}\cup X\in\mathcal{C}(\mathcal{E})$, then $C\xRightarrow{X} C'$ is called a weak pomset transition from $C$ to $C'$, where we define $\xRightarrow{e}\triangleq\xrightarrow{\tau^*}\xrightarrow{e}\xrightarrow{\tau^*}$. And $\xRightarrow{X}\triangleq\xrightarrow{\tau^*}\xrightarrow{e}\xrightarrow{\tau^*}$, for every $e\in X$. When the events in $X$ are pairwise concurrent, we say that $C\xRightarrow{X}C'$ is a weak step.

We will also suppose that all the PESs in this paper are image finite, that is, for any PES $\mathcal{E}$ and $C\in \mathcal{C}(\mathcal{E})$ and $a\in \Lambda$, $\{e\in\hat{E}|C\xRightarrow{e} C'\wedge \lambda(e)=a\}$ is finite.

\subsection{Weakly Concurrent Behavioral Equivalence}

\textbf{Definition \ref{wtc}.4 (Weak Pomset, Step Bisimulation).} Let $\mathcal{E}_1$, $\mathcal{E}_2$ be PESs. A weak pomset bisimulation is a relation $R\subseteq\mathcal{C}(\mathcal{E}_1)\times\mathcal{C}(\mathcal{E}_2)$, such that if $(C_1,C_2)\in R$, and $C_1\xRightarrow{X_1}C_1'$ then $C_2\xRightarrow{X_2}C_2'$, with $X_1\subseteq \hat{E_1}$, $X_2\subseteq \hat{E_2}$, $X_1\sim X_2$ and $(C_1',C_2')\in R$, and vice-versa. We say that $\mathcal{E}_1$, $\mathcal{E}_2$ are weak pomset bisimilar, written $\mathcal{E}_1\approx_p\mathcal{E}_2$, if there exists a weak pomset bisimulation $R$, such that $(\emptyset,\emptyset)\in R$. By replacing weak pomset transitions with weak steps, we can get the definition of weak step bisimulation. When PESs $\mathcal{E}_1$ and $\mathcal{E}_2$ are weak step bisimilar, we write $\mathcal{E}_1\approx_s\mathcal{E}_2$.

\textbf{Definition \ref{wtc}.5 (Posetal Product).} Given two PESs $\mathcal{E}_1$, $\mathcal{E}_2$, the posetal product of their configurations, denoted $\mathcal{C}(\mathcal{E}_1)\overline{\times}\mathcal{C}(\mathcal{E}_2)$, is defined as

$$\{(C_1,f,C_2)|C_1\in\mathcal{C}(\mathcal{E}_1),C_2\in\mathcal{C}(\mathcal{E}_2),f:\hat{C_1}\rightarrow \hat{C_2} \textrm{ isomorphism}\}.$$

A subset $R\subseteq\mathcal{C}(\mathcal{E}_1)\overline{\times}\mathcal{C}(\mathcal{E}_2)$ is called a posetal relation. We say that $R$ is downward closed when for any $(C_1,f,C_2),(C_1',f,C_2')\in \mathcal{C}(\mathcal{E}_1)\overline{\times}\mathcal{C}(\mathcal{E}_2)$, if $(C_1,f,C_2)\subseteq (C_1',f',C_2')$ pointwise and $(C_1',f',C_2')\in R$, then $(C_1,f,C_2)\in R$.

For $f:X_1\rightarrow X_2$, we define $f[x_1\mapsto x_2]:X_1\cup\{x_1\}\rightarrow X_2\cup\{x_2\}$, $z\in X_1\cup\{x_1\}$,(1)$f[x_1\mapsto x_2](z)=
x_2$,if $z=x_1$;(2)$f[x_1\mapsto x_2](z)=f(z)$, otherwise. Where $X_1\subseteq \hat{E_1}$, $X_2\subseteq \hat{E_2}$, $x_1\in \hat{E}_1$, $x_2\in \hat{E}_2$.

\textbf{Definition \ref{wtc}.6 (Weak (Hereditary) History-Preserving Bisimulation).} A weak history-preserving (hp-) bisimulation is a posetal relation $R\subseteq\mathcal{C}(\mathcal{E}_1)\overline{\times}\mathcal{C}(\mathcal{E}_2)$ such that if $(C_1,f,C_2)\in R$, and $C_1\xRightarrow{e_1} C_1'$, then $C_2\xRightarrow{e_2} C_2'$, with $(C_1',f[e_1\mapsto e_2],C_2')\in R$, and vice-versa. $\mathcal{E}_1,\mathcal{E}_2$ are weak history-preserving (hp-)bisimilar and are written $\mathcal{E}_1\sim_{hp}\mathcal{E}_2$ if there exists a hp-bisimulation $R$ such that $(\emptyset,\emptyset,\emptyset)\in R$.

A weak hereditary history-preserving (hhp-)bisimulation is a downward closed weak hp-bisimulation. $\mathcal{E}_1,\mathcal{E}_2$ are weak hereditary history-preserving (hhp-)bisimilar and are written $\mathcal{E}_1\approx_{hhp}\mathcal{E}_2$.

\textbf{Proposition \ref{wtc}.7 (Weakly Concurrent Behavioral Equivalence).} (Strongly) concurrent behavioral equivalences imply weakly concurrent behavioral equivalences. That is, $\sim_p$ implies $\approx_p$, $\sim_s$ implies $\approx_s$, $\sim_{hp}$ implies $\approx_{hp}$, $\sim_{hhp}$ implies $\approx_{hhp}$.

\begin{proof}
From the definition of weak pomset transition, weak step transition, posetal product and weakly concurrent behavioral equivalence, it is easy to see that $\xrightarrow{e}=\xrightarrow{\epsilon}\xrightarrow{e}\xrightarrow{\epsilon}$ for $e\in E$, where $\epsilon$ is the empty event.
\end{proof}

\section{A Logic for Weakly True Concurrency}{\label{logic}}

In this section, we will introduce a logic for weakly true concurrency, which is called $\mathcal{L}^{\tau}$, and the logic equivalence induced by the logic.

\subsection{A Logic $\mathcal{L}^{\tau}$}

Let $\textbf{x}$ and $\textbf{y}$ denote tuples of variables $x_1,\cdots,x_n$ and $y_1,\cdots,y_n$. Next, we give the syntax of the logic $\mathcal{L}^{\tau}$.

\textbf{Definition \ref{logic}.1 (Syntax).} Let $Var$ be a denumerable set of variables ranged over by $x,y,z,\cdots$. The syntax of the logic $\mathcal{L}^{\tau}$ over the set of labels $\Lambda$ is defined as follows, where $a$ ranges over $\Lambda$:

$$\varphi ::= \textsf{T} | \varphi\wedge\varphi | \neg\varphi | (\textbf{x},\overline{\textbf{y}}\ll az)\varphi | \langle\langle z\rangle\rangle \varphi$$

The notions of free variables, environments, and legal pairs are same as those of the logic $\mathcal{L}$ for (strongly) true concurrency \cite{LTC1} \cite{LTC2}, but which are only consider events in $\hat{E}$. We retype these concepts as follows.

The set of free variables of a formula $\varphi$, $fv(\varphi)$, is defined inductively as follows.

\begin{enumerate}
  \item $fv(\textsf{T})=\emptyset$
  \item $fv(\varphi_1\wedge\varphi_2)=fv(\varphi_1)\cup fv(\varphi_2)$
  \item $fv(\neg\varphi)=fv(\varphi)$
  \item $fv((\textbf{x},\overline{\textbf{y}}\ll az)\varphi)=\textbf{x}\cup\textbf{y}\cup(fv(\varphi)\backslash\{z\})$
  \item $fv(\langle\langle z\rangle\rangle)\varphi=fv(\varphi)\cup\{z\}$
\end{enumerate}

Let $\mathcal{E}$ be a PES, and let $Env_{\mathcal{E}}$ be the set of environments $\eta:Var\rightarrow \hat{E}$. Given a formula $\varphi$ in $\mathcal{L}^{\tau}$, a pair $(C,\eta)\in\mathcal{C}(\mathcal{E})\times Env_{\mathcal{E}}$ is legal for $\varphi$ if $C\cup\eta(fv(\varphi))\cup\{\tau|\tau^*e\tau^*, e\in C\cup\eta(fv(\varphi))\}$ is a consistent set of events. We denote by $lp_{\mathcal{E}}(\varphi)$ the set of legal pairs for $\varphi$ in $\mathcal{E}$. And let $E[C]=\{e|e\in \hat{E}\backslash \hat{C} \wedge C\frown e\}$ called residual of $E$ after $C$.

\textbf{Definition \ref{logic}.2 (Semantics).} Let $\mathcal{E}$ be a PES. The denotation of a formula $\varphi$, denoted $\{|\varphi|\}^{\mathcal{E}}\in 2^{\mathcal{C}(\mathcal{E})\times Env_{\mathcal{E}}}$, is defined inductively as follows:

\begin{enumerate}
  \item $\{|\textsf{T}|\}^{\mathcal{E}}=\mathcal{C}(\mathcal{E}\times Env_{\mathcal{E}})$
  \item $\{|\varphi_1\wedge\varphi_2|\}^{\mathcal{E}}=\{|\varphi_1|\}^{\mathcal{E}}\cap\{|\varphi_2|\}^{\mathcal{E}}\cap lp(\varphi_1\wedge\varphi_2)$
  \item $\{|\neg\varphi|\}^{\mathcal{E}}=lp(\varphi)\backslash\{|\varphi|\}$
  \item $\{|(\textbf{x},\overline{\textbf{y}}\ll az)\varphi|\}^{\mathcal{E}}=\{(C,\eta)|(C,\eta)\in lp((\textbf{x},\overline{\textbf{y}}\ll az)\varphi)\textrm{ and } \exists e\in\hat{E(C)}\textrm{ such that } e\frown \eta(fv(\varphi)\backslash\{z\})\wedge \lambda(e)=a \wedge \eta(\textbf{x})\ll e \wedge \eta(\textbf{y})\parallel(\tau^* e \tau^*)\wedge (C,\eta[z\mapsto e])\in \{|\varphi|\}^{\mathcal{E}}\}$
  \item $\{|\langle\langle z\rangle\rangle\varphi|\}^{\mathcal{E}}=\{(C,\eta)|C\xRightarrow{\eta(z)} C' \wedge (C',\eta)\in \{|\varphi|\}^{\mathcal{E}}\}$
\end{enumerate}

Also, the PES $\mathcal{E}$ satisfies the formula $\varphi$ in the configuration $C$ and the environment $\eta$. when $(C,\eta)\in \{|\varphi|\}^{\mathcal{E}}$, written $\mathcal{E},C\models_{\eta}\varphi$. For closed formulae $\varphi$, we write $\mathcal{E},C\models\varphi$, when $\mathcal{E},C\models_{\eta}\varphi$ for some $\eta$. And $\mathcal{E}\models\varphi$, when $\mathcal{E},\emptyset\models\varphi$.

There are two formula $(\textbf{x},\overline{\textbf{y}}\ll az)\varphi$ and $\langle\langle z\rangle\rangle \varphi$ in $\mathcal{L}^{\tau}$, different to $(\textbf{x},\overline{\textbf{y}}< az)\varphi$ and $\langle z\rangle \varphi$ in $\mathcal{L}$. The difference between $(\textbf{x},\overline{\textbf{y}}\ll az)\varphi$ and $(\textbf{x},\overline{\textbf{y}}< az)\varphi$ is that the event $e$ binding to $z$ is caused by the events already bound to variables in $\textbf{x}$ and also some silent events $\tau^*$, and $e$ binding to $z$ is independent from those bound to variables in $\textbf{y}$, and each event bound to some variable in $\textbf{y}$ may follow some $\tau$ events and may have some $\tau$ events followed. Also, the difference between $\langle\langle z\rangle\rangle \varphi$ and $\langle z\rangle \varphi$ in $\mathcal{L}$ is that $e$ binding to $z$ is executed, and may follow some $\tau$ events and may have some $\tau$ events followed.

Given a PES $\mathcal{E}$, free variables in a formula $\varphi$, environments $\eta:Var\rightarrow \hat{E}$, and a legal pair $(C,\eta)\in\mathcal{C}(\mathcal{E})\times Env_{\mathcal{E}}$ only consider events $\hat{E}$, and ensure that every $\tau$ event keeps silent. Several conclusions on legal pairs and environments in $\mathcal{L}$ will still hold in $\mathcal{L}^{\tau}$. We only retype the conclusions without proofs.

\textbf{Lemma \ref{logic}.3 (Negation).} Let $\varphi$ be a closed formula in $\mathcal{L^{\tau}}$, let $\mathcal{E}$ be a PES and let $(C,\eta)\in \mathcal{C}(\mathcal{E})\times Env_{\mathcal{E}}$. Then, $\mathcal{E},C\models_{\eta}\varphi$ iff $\mathcal{E},C\nvDash_{\eta}\neg\varphi$.

\textbf{Lemma \ref{logic}.4 (Denotations Consist of Legal Pairs).} Let $\mathcal{E}$ be a PES. Then, for any formula $\varphi\in\mathcal{L}^{\tau}$, it holds $\{|\varphi|\}^{\mathcal{E}}\subseteq lp_{\mathcal{E}}(\varphi)$.

\textbf{Lemma \ref{logic}.5.} Let $\mathcal{E}$ be a PES and let $C\in\mathcal{C}(\mathcal{E})$. Let $\varphi\in\mathcal{L}^{\tau}$ and let $\eta_1,\eta_2:Var\rightarrow \hat{E}$ be environments such that $\eta_1(x)=\eta_2(x)$ for any $x\in fv(\varphi)$. Then, $\mathcal{E},C\models_{\eta_1}\varphi$ iff $\mathcal{E},C\models_{\eta_2}\varphi$.

For dual operators, the new cases in $\mathcal{L}^{\tau}$ are as follows.

$\{\textbf{x},\overline{\textbf{y}}\ll az\}\varphi$ for the formula $\neg((\textbf{x},\overline{\textbf{y}}\ll az)\neg\varphi)$. Its semantics is as follows.

$\{|\{\textbf{x},\overline{\textbf{y}}\ll az\}\varphi|\}^{\mathcal{E}}=\{(C,\eta)|(C,\eta)\in lp(\{\textbf{x},\overline{\textbf{y}}\ll az\}\varphi) \textrm{ and }\forall e\in\hat{E(C)} \textrm{ such that } e\frown \eta(fv(\varphi)\backslash\{z\})\wedge \lambda(e)=a \wedge \eta(\textbf{x})\ll e \wedge \eta(\textbf{y})\parallel(\tau^* e \tau^*)\wedge (C,\eta[z\mapsto e])\in \{|\varphi|\}^{\mathcal{E}}\}$.

$[[\varphi]]$ for the formula $\neg(\langle\langle z\rangle\rangle\neg\varphi)$. Its semantics is as follows.

$\{|[[z]]|\}^{\mathcal{E}}=\{(C,\eta)|(C,\eta)\in lp([[z]]\varphi) \textrm{ and if } C\xRightarrow{\eta(z)} C' \textrm{ then } (C',\eta)\in \{|\varphi|\}^{\mathcal{E}}\}$.

We also write $\langle\langle |\textbf{x},\overline{\textbf{y}}\ll az |\rangle\rangle\varphi$ for the formula $(\textbf{x},\overline{\textbf{y}}\ll az)\langle\langle z\rangle\rangle\varphi$.

And $((\textbf{x},\overline{\textbf{y}}\ll az)\otimes(\textbf{x}',\overline{\textbf{y}'}\ll bz'))\varphi$ for the formula $(\textbf{x},\overline{\textbf{y}}\ll az)(\textbf{x}',\overline{\textbf{y}',z}\ll bz')\varphi$, and $(\langle\langle |\textbf{x},\overline{\textbf{y}}\ll az|\rangle\rangle\otimes\langle\langle |\textbf{x}',\overline{\textbf{y}'}\ll bz'|\rangle\rangle)\varphi$ for the formula $(\textbf{x},\overline{\textbf{y}}\ll az)\otimes(\textbf{x}',\overline{\textbf{y}'}\ll bz')\langle\langle z\rangle\rangle\langle\langle z'\rangle\rangle\varphi$.

\subsection{The Logic $\mathcal{L}^{\tau}$ and Weak HHP-Bisimilarity $\approx_{hhp}$}

We will show that the logic $\mathcal{L}^{\tau}$ induces weak hhp-bisimilarity.

Two PESs $\mathcal{E}_1$, $\mathcal{E}_2$ are logically equivalent in $\mathcal{L}^{\tau}$, written $\mathcal{E}_1\equiv_{\mathcal{L}^{\tau}}\mathcal{E}_2$, when they satisfy the same closed formulae in $\mathcal{L}^{\tau}$.

\textbf{Proposition \ref{logic}.6.} Let $\mathcal{E}_1$ and $\mathcal{E}_2$ be PESs such that $\mathcal{E}_1\equiv_{\mathcal{L}^{\tau}}\mathcal{E}_2$, then $\mathcal{E}_1\approx_{hhp}\mathcal{E}_2$.

\begin{proof}
We should remember the difference between $\mathcal{L}^{\tau}$ and $\mathcal{L}$, also $\approx_{hhp}$ and $\sim_{hhp}$. The basic difference is that in $\mathcal{L}^{\tau}$ and $\approx_{hhp}$, the execution of an event $e$ may follow some $\tau$ events and may have some $\tau$ events followed.

So the proof is similar with the soundness proof $\mathcal{E}_1\equiv_{\mathcal{L}}\mathcal{E}_2$, implies $\mathcal{E}_1\sim_{hhp}\mathcal{E}_2$. We just give the skeleton of the proof.

We fix a surjective environment $\eta_1:Var\rightarrow\hat{E_1}$. For an event $e\in\hat{E_1}$, $\eta_1(x_e)=e$. For a configuration $\hat{C_1}=\{e_1,\cdots e_n\}$, the set of variables $X_{\hat{C_1}}=\{x_{e_1},\cdots,x_{e_n}\}$.

Consider the posetal relation $R\subseteq\mathcal{C}(\mathcal{E}_1)\overline{\times}\mathcal{C}(\mathcal{E}_2)$ defined by:

\begin{equation}
R=\{(C_1,f,C_2)|\forall \psi\in\mathcal{L}^{\tau}.fv(\psi)\subseteq X_{\hat{C_1}} (\mathcal{E}_1,\emptyset\models_{\eta_1}\psi\textrm{ iff }\mathcal{E}_2,\emptyset\models_{f\circ\eta_1}\psi)\}.
\end{equation}

Since $\mathcal{E}_1\equiv_{\mathcal{L}^{\tau}}\mathcal{E}_2$, we have that $(\emptyset,\emptyset,\emptyset)\in R$. It is still sufficient to prove that $R$ is a weak hhp-bisimulation $\approx_{hhp}$.

Since in $\mathcal{L}^{\tau}$ and $\approx_{hhp}$, we only consider the events in $\hat{E_1}$ and $\hat{E_2}$, it is easy to see that $R$
in (1) is downward closed.

To prove $R$ is a weak hp-bisimulation, it is sufficient to show that given $(C_1,f,C_2)\in R$, if $C_1\xRightarrow{e}C_1'$, then there exists a transition $C_2\xRightarrow{g}C_2'$, such that $f'=f[e\mapsto g]:\hat{C_1'}\rightarrow \hat{C_2'}$ is an isomorphism of pomsets and $(C_1',f',C_2')\in R$.

Since all PESs are assumed to be image finite, there are finitely many transitions $C_2\xRightarrow{g^i} C_2^i$ ($i\in\{1,\cdots,n\}$), such that $\hat{C_1'}\sim \hat{C_2^i}$ as pomset.

Then we proceed by contradiction. Assume that, for any $i\in\{1,\cdots,n\}$, it holds $(C_1',f^i,C_2^i)\notin R$. By the definition of $R$ (1), there exists a formula $\psi^i$ such that $\mathcal{E}_1,\emptyset\models_{\eta_1}\psi^i$ and $\mathcal{E}_2,\emptyset\nvDash_{f^i\circ\eta_1}\psi^i$, where $fv(\psi^i)\subseteq X_{\hat{C_1'}}=X_{\hat{C_1}}\cup\{x_e\}$ and $f^i=f[e\mapsto g^i]$.

Similarly, constructing the following formula

$$\varphi=(\textbf{x},\overline{\textbf{y}}\ll ax_e)(\langle\langle X_{\hat{C_1}}\rangle\rangle\langle\langle x_e\rangle\rangle\textsf{T}\wedge\psi^1\wedge\cdots\wedge\psi^n)$$,

where $a=\lambda_1(e)$ and $\textbf{x},\textbf{y}\subseteq X_{\hat{C_1}}$.

It can be proven that $\mathcal{E}_1,\emptyset\models_{\eta_1}\varphi$, and $\mathcal{E}_2,\emptyset\nvDash_{f\circ\eta_1}\varphi$, based on the contradiction hypothesis.

The converse also can be proven analogously.
\end{proof}

The lemma about hhp-bisimilarity as a PES \cite{HHP1} still holds for the case of the weak hhp-bisimilarity, that is, weak hhp-bisimilarity can also act as a PES. This leads to the following lemma.

\textbf{Lemma \ref{logic}.7 (Weak Hhp-Bisimilarity as a PES).} Let $\mathcal{E}_1$, $\mathcal{E}_2$ be PESs such that $\mathcal{E}_1\approx_{hhp}\mathcal{E}_2$, and let $R$ be a weak hhp-bisimulation. Then there exists a PES $\mathcal{E}_R=\langle E_R,\leq_R,\sharp_R,\lambda_R\rangle$ such that for $i\in\{1,2\}$:

\begin{enumerate}
  \item $\mathcal{E}_i\approx_{hhp}\mathcal{E}_R$;
  \item there are surjective maps $f^i_R: \hat{E_R}\rightarrow \hat{E_i}$ such that $\{(C,f^i_{R|C},f^i_R(\hat{C}))|C\in\mathcal{C}(\mathcal{E}_R)\}$ is a weak hhp-bisimulation.
\end{enumerate}

Additionally, each $f^i_R$ preserves labels, causality $\leq$ and concurrency $\parallel$, it maps configurations to configurations and it is injective on consistent sets of events.

\begin{proof}
We just restrict the event $e_1$, $e_2$, $e_1'$, $e_2'$ in the sets $\hat{E_1}$ and $\hat{E_2}$, then we can construct $\mathcal{E}_R=\langle E_R,\leq_R,\sharp_R,\lambda_R\rangle$:

\begin{enumerate}
  \item $E_R=\{(e_1,f,e_2)|(\ulcorner e_1\urcorner,f,\ulcorner e_2\urcorner)\in R\}$,
  \item $(e_1,f,e_2)\leq_R(e_1',f',e_2')$ if $f\subseteq f'$,
  \item $(e_1,f,e_2)\sharp_R(e_1',f',e_2')$ if there exists no $(C,g,D)\in R$ such that $(\ulcorner e_1\urcorner,f,\ulcorner e_2\urcorner),(\ulcorner e_1'\urcorner,f,\ulcorner e_2'\urcorner)\subseteq (C,g,D)$ pointwise,
  \item $\lambda_R(e_1,f,e_2)=\lambda_1(e_1)$.
\end{enumerate}

The maps $f^1_R: \hat{E_R}\rightarrow \hat{E_1}$ and $f^2_R: \hat{E_R}\rightarrow \hat{E_2}$ are just the projections on the first and third components, respectively.
\end{proof}

\textbf{Proposition \ref{logic}.8.} Let $\mathcal{E}_1$ and $\mathcal{E}_2$ are PESs, if $\mathcal{E}_1\approx_{hhp}\mathcal{E}_2$, then $\mathcal{E}_1\equiv_{\mathcal{L}^{\tau}}\mathcal{E}_2$.

\begin{proof}
Let $R$ be a weak hhp-bisimulation relating $\mathcal{E}_1$ and $\mathcal{E}_2$. We assume that $R=\{(C_1,f_{|\hat{C_1}},f(\hat{C_1}))\}$, where $f: \hat{E_1}\rightarrow\hat{E_2}$ is a surjective map satisfying the condition of Lemma \ref{logic}.7 and preserves legal pairs. It is still sufficient to prove that for any formula $\varphi\in\mathcal{L}^{\tau}$, and any $(C_1,\eta_1)\in lp_{\mathcal{E}_1}(\varphi)$

\begin{equation}
\mathcal{E}_1,C_1\models_{\eta_1}\varphi \textrm{ iff } \mathcal{E}_2,f(\hat{C_1})\models_{f\circ\eta_1}\varphi
\end{equation}

From (2), we can get $\mathcal{E}_1\models \varphi$ iff $\mathcal{E}_2\models \varphi$, for any closed formula $\varphi$. That is, $\mathcal{E}_1\equiv_{\mathcal{L}^{\tau}}\mathcal{E}_2$.

The proof proceeds by induction on the formula $\varphi$. That is, the cases include $\varphi=\textsf{T}$, $\varphi=\varphi_1\wedge\varphi_2$, $\varphi=\neg\varphi_1$,$\varphi=(\textbf{x},\overline{\textbf{y}}\ll az)\psi$, and $\varphi=\langle\langle x\rangle\rangle\psi$. We only prove the non-trivial cases $\varphi=(\textbf{x},\overline{\textbf{y}}\ll az)\psi$ and $\varphi=\langle\langle x\rangle\rangle\psi$.

The cases $\varphi=(\textbf{x},\overline{\textbf{y}}\ll az)\psi$ and $\varphi=\langle\langle x\rangle\rangle\psi$ are as follows.

We notice that $f$ in (2) is a map from $\hat{E_1}$ to $\hat{E_2}$, the definitions of free variables, semantics and legal pairs in $\varphi\in\mathcal{L}^{\tau}$ make $\tau$ keeps silent. So, by the definitions of semantics for $\varphi=(\textbf{x},\overline{\textbf{y}}\ll az)\psi$ and $\varphi=\langle\langle x\rangle\rangle\psi$, Lemma \ref{logic}.4, and Lemma \ref{logic}.7, $\mathcal{E}_1,C_1\models_{\eta_1}\varphi \textrm{ iff } \mathcal{E}_2,f(\hat{C_1})\models_{f\circ\eta_1}\varphi$ with $\mathcal{E}_1\approx_{hhp}\mathcal{E}_2$, for $\varphi=(\textbf{x},\overline{\textbf{y}}\ll az)\psi$ and $\varphi=\langle\langle x\rangle\rangle\psi$ in $\mathcal{L}^{\tau}$, can be proven similarly to $\mathcal{E}_1,C_1\models_{\eta_1}\varphi \textrm{ iff } \mathcal{E}_2,f(C_1)\models_{f\circ\eta_1}\varphi$ with $\mathcal{E}_1\sim_{hhp}\mathcal{E}_2$, for $\varphi=(\textbf{x},\overline{\textbf{y}}< az)\psi$ and $\varphi=\langle x\rangle\psi$ in $\mathcal{L}$. We do not retype the proof processes.
\end{proof}

Proposition \ref{logic}.6 and \ref{logic}.8 together say that weak hhp-bisimilarity is the logical equivalence of $\mathcal{L}^{\tau}$.

\textbf{Theorem \ref{logic}.9 (Weak Hhp-Bisimilarity, Logically).} Let $\mathcal{E}_1$ and $\mathcal{E}_2$ be PESs. Then, $\mathcal{E}_1\approx_{hhp}\mathcal{E}_2$ iff $\mathcal{E}_1\equiv_{\mathcal{L}^{\tau}}\mathcal{E}_2$.

\section{Fragments of the Logic $\mathcal{L}^{\tau}$}\label{frag}

It is natural to find that different fragments of the logic $\mathcal{L}^{\tau}$ induce corresponding concurrent equivalences: Hennessy-Milner logic $\mathcal{L}^{\tau}_{HM}$ induces weak (interleaving) bisimilarity $\approx_{HM}$, step logic $\mathcal{L}^{\tau}_{s}$ induces weak step bisimilarity $\approx_s$, pomset logic $\mathcal{L}^{\tau}_{p}$ induces weak pomset bisimilarity $\approx_p$, and hp logic $\mathcal{L}^{\tau}_{hp}$ induces weak hp-bisimilarity $\approx_{hp}$.

\subsection{Hennessy-Milner Logic $\mathcal{L}^{\tau}_{HM}$ and Weak (Interleaving) Bisimilarity $\approx_{HM}$}

The syntax of Hennessy-Milner logic $\mathcal{L}^{\tau}_{HM}$ is as follows:

\begin{center}
$$\varphi::=\langle\langle |ax|\rangle\rangle\varphi | \varphi\wedge\varphi | \neg\varphi | \textsf{T}$$
\end{center}

It is well known that equivalence $\mathcal{L}^{\tau}_{HM}$ induced is weak (interleaving) bisimilarity \cite{ALNC} \cite{CC} as follows.

\textbf{Theorem \ref{frag}.1 (Weak (Interleaving) Bisimilarity, Logically).} Let $\mathcal{E}_1$ and $\mathcal{E}_2$ be PESs. Then, $\mathcal{E}_1\approx_{HM}\mathcal{E}_2$ iff $\mathcal{E}_1\equiv_{\mathcal{L}^{\tau}_{HM}}\mathcal{E}_2$.

\subsection{Step Logic $\mathcal{L}^{\tau}_{s}$ and Weak Step Bisimilarity $\approx_{s}$}

The syntax of step logic $\mathcal{L}^{\tau}_{s}$ is as follows:

\begin{center}
$$\varphi::=(\langle\langle |a_1 x_1|\rangle\rangle\otimes\cdots\otimes\langle\langle |a_n x_n| \rangle\rangle)\varphi | \varphi\wedge\varphi | \neg\varphi | \textsf{T}$$
\end{center}

\textbf{Theorem \ref{frag}.2 (Weak Step Bisimilarity, Logically).} Let $\mathcal{E}_1$ and $\mathcal{E}_2$ be PESs. Then, $\mathcal{E}_1\approx_s\mathcal{E}_2$ iff $\mathcal{E}_1\equiv_{\mathcal{L}^{\tau}_{s}}\mathcal{E}_2$.

\begin{proof}
All formulae in $\mathcal{L}^{\tau}_s$ are closed, let $C_1\in\mathcal{C}(\mathcal{E}_1)$ and $C_2\in\mathcal{C}(\mathcal{E}_2)$, and let $R$ be a weak step bisimulation, we need to prove that $(C_1,C_2)\in R$, iff for any formula $\varphi\in\mathcal{L}^{\tau}_s$, $\mathcal{E}_1,C_1\models\varphi\Leftrightarrow\mathcal{E}_2,C_2\models\varphi$.

($\Rightarrow$)For $(C_1,C_2)\in R$ and $R$ is a weak step bisimulation, to get $\mathcal{E}_1,C_1\models\varphi\Leftrightarrow\mathcal{E}_2,C_2\models\varphi$ for all $\varphi\in\mathcal{L}^{\tau}_s$, we need to induct on the structure of $\varphi$.

For the nontrivial case $\varphi=(\langle\langle |a_1 x_1|\rangle\rangle\otimes\cdots\otimes\langle\langle |a_n x_n| \rangle\rangle)\psi$, all events $\{e_1,\cdots, e_n\}\subseteq \hat{E_1}$, a weak step $C_1\xRightarrow{\{e_1,\cdots, e_n\}}C_1'$ will cause a weak step $C_2\xRightarrow{\{g_1,\cdots, g_n\}}C_2'$ in which $\{g_1,\cdots, g_n\}\subseteq \hat{E_2}$. And each event $e_i\in \{e_1,\cdots, e_n\}$, $g_i\in \{g_1,\cdots, g_n\}$ may follow $\tau^*$ and may have $\tau^*$ followed.

$(C_1',C_2')\in R$, $\mathcal{E}_1,C_1\models \varphi$, $\mathcal{E}_1,C_1'\models_{\eta_1[x_1\mapsto e_1,\cdots,x_n\mapsto e_n]}\psi$, $\mathcal{E}_2,C_2'\models_{\eta_2[x_1\mapsto g_1,\cdots,x_n\mapsto g_n]}\psi$, $\mathcal{E}_2,C_2\models_{\eta_2} \varphi$, since $\varphi$ is closed, $\mathcal{E}_2,C_2\models \varphi$.

($\Leftarrow$)We prove that the relation

\begin{center}
    $$R=\{(C_1,C_2)|\forall\varphi\in\mathcal{L}^{\tau}_s (\mathcal{E}_1,C_1\models\varphi\textrm{ iff }\mathcal{E}_2,C_2\models\varphi)\}$$
\end{center}

is a weak step bisimulation.

We proceed by contradiction. For $X\subseteq\hat{E_1}$ and $Y\subseteq\hat{E_2}$, let $(C_1,C_2)\in R$, $C_1\xRightarrow{X}C_1'$, $C_2\xRightarrow{Y}C_2'$ and $X\sim Y$ as pomsets, then $(C_1',C_2'\notin R)$. Hence, there exists a formula $\psi\in\mathcal{L}^{\tau}_s$, such that $\mathcal{E}_1,C_1'\models\psi$ and $\mathcal{E}_2,C_2'\nvDash\psi$.

Since all PESs are image finite, we assume some step $Y^i$ and some formula $\psi^i$, $\mathcal{E}_1,C_1'\models\psi^i$ and $\mathcal{E}_2,C_2'\nvDash\psi^i$.

Then we can construct a formula

\begin{center}
    $$\varphi=(\langle\langle |a_1 x_1|\rangle\rangle\otimes\cdots\otimes\langle\langle |a_n x_n| \rangle\rangle)(\psi^1\wedge\cdots\wedge\psi^k)$$,
\end{center}

such that $\mathcal{E}_1,C_1\models\varphi$ and $\mathcal{E}_2,C_2\nvDash\varphi$.
\end{proof}

\subsection{Pomset Logic $\mathcal{L}^{\tau}_{p}$ and Weak Pomset Bisimilarity $\approx_{p}$}

The syntax of pomset logic $\mathcal{L}^{\tau}_{p}$ is as follows:

\begin{center}
$$\varphi::=\langle\langle |\textbf{x},\overline{\textbf{y}}\ll az|\rangle\rangle\varphi | \varphi\wedge\varphi | \neg\varphi | \textsf{T}$$
where $\neg$ and $\wedge$ are used only on closed formulae.
\end{center}

Let $Pom(\langle\langle | \textbf{x}_1,\overline{\textbf{y}_1}\ll a_1 z_1|\rangle\rangle\cdots\langle\langle | \textbf{x}_n,\overline{\textbf{y}_n}\ll a_n z_n|\rangle\rangle)$, denote the class of pomsets $(Z,\leq,\lambda)$ such that $Z=\{z_1,\cdots z_n\}$ and $\lambda(z_i)=a_i$, and for any $z\in Z$, (1)$z\in \textbf{x}_i$ implies $z\leq z_i$, (2)$z\in\textbf{y}_i$ implies $z\nleq z_i$.

The following lemma still stands for weak pomset transition.

\textbf{Lemma \ref{frag}.3.} Let $\varphi=(\langle\langle | \textbf{x}_1,\overline{\textbf{y}_1}\ll a_1 z_1|\rangle\rangle\cdots\langle\langle | \textbf{x}_n,\overline{\textbf{y}_n}\ll a_n z_n|\rangle\rangle)\psi$ be a closed formula in $\mathcal{L}^{\tau}_{p}$. Then $\mathcal{E},C\models_{\eta}\varphi$ iff $C\xRightarrow{X}C'$ where $X=\{e_1,\cdots,e_n\}$ is a pomset such that $X\sim(Z,\leq,\lambda)$ for some $(Z,\leq,\lambda)\in Pom(\langle\langle | \textbf{x}_1,\overline{\textbf{y}_1}\ll a_1 z_1|\rangle\rangle\cdots\langle\langle | \textbf{x}_n,\overline{\textbf{y}_n}\ll a_n z_n|\rangle\rangle)$ and $\mathcal{E},C'\models_{\eta'}\psi$, with $\eta'=\eta[z_1\mapsto e_1,\cdots,z_n\mapsto e_n]$.

Since $\tau$ in the execution of a single pomset keeps silent, the execution of a single pomset still can be characterized by a corresponding formula in $\mathcal{L}^{\tau}_p$. That is, for $p_Z=(Z,\leq_{p_Z},\lambda_{p_Z})$, $\varphi\in \mathcal{L}^{\tau}_p$, we define $\langle\langle |p_Z| \rangle\rangle$ as follows: (1) $\langle\langle |p_Z| \rangle\rangle\varphi= \varphi  \textrm{ if } Z=\emptyset$; (2)$\langle\langle |p_Z| \rangle\rangle\varphi=\langle\langle |p_{Z'}| \rangle\rangle\langle\langle| \textbf{x},\overline{\textbf{y}}\ll az |\rangle\rangle\varphi$, where $Z=Z'\cup\{z\}$, $z$ is maximal with respect to $\leq_{p_Z}$, $\textbf{x}=\{z'\in Z'|z'\leq_{p_Z}z\}$, $\textbf{y}=Z'\backslash\textbf{x}$, $a=\lambda_{p_Z}(z)$.

Lemma \ref{frag}.3 holds for the above pomset formula. Let $\mathcal{E}$ be a PES and $C\in\mathcal{C}(\mathcal{E})$, $\{z_1,\cdots,z_n\}\subseteq Var$, $p_Z=(Z,\leq_{p_Z},\lambda_{p_Z})$, then $\mathcal{E},C\models\langle\langle|p_Z|\rangle\rangle\varphi$ iff $C\xRightarrow{X}C'$ where $X=\{e_1,\cdots,e_n\}$ is a pomset and $X\sim p_Z$ and $\mathcal{E},C'\models_{\eta'}\varphi$, with $\eta'=\eta[z_1\mapsto e_1,\cdots,z_n\mapsto e_n]$.

\textbf{Theorem \ref{frag}.4 (Weak Pomset Bisimilarity, Logically)} Let $\mathcal{E}_1$ and $\mathcal{E}_2$ be PESs. Then, $\mathcal{E}_1\approx_p\mathcal{E}_2$ iff $\mathcal{E}_1\equiv_{\mathcal{L}^{\tau}_{p}}\mathcal{E}_2$.

\begin{proof}
Let $C_1\in\mathcal{C}(\mathcal{E}_1)$ and $C_2\in\mathcal{C}(\mathcal{E}_2)$, and let $R$ be a weak pomset bisimulation, we need to prove that $(C_1,C_2)\in R$, iff for any formula $\varphi\in\mathcal{L}^{\tau}_p$, $\mathcal{E}_1,C_1\models\varphi\Leftrightarrow\mathcal{E}_2,C_2\models\varphi$.

($\Rightarrow$)For $(C_1,C_2)\in R$ and $R$ is a weak pomset bisimulation, to get $\mathcal{E}_1,C_1\models\varphi\Leftrightarrow\mathcal{E}_2,C_2\models\varphi$ for all $\varphi\in\mathcal{L}^{\tau}_p$, we need to induct on the structure of $\varphi$.

For the nontrivial case $\varphi=(\langle\langle |\textbf{x}_1,\overline{\textbf{y}_1}\ll a_1 z_1|\rangle\rangle\cdots\langle\langle |\textbf{x}_n,\overline{\textbf{y}_n}\ll a_n z_n| \rangle\rangle)\psi$, all events $\{e_1,\cdots, e_n\}\subseteq \hat{E_1}$, a weak pomset transition $C_1\xRightarrow{\{e_1,\cdots, e_n\}}C_1'$ will cause a weak pomset $C_2\xRightarrow{\{g_1,\cdots, g_n\}}C_2'$ in which $\{g_1,\cdots, g_n\}\subseteq \hat{E_2}$. And each event $e_i\in \{e_1,\cdots, e_n\}$, $g_i\in \{g_1,\cdots, g_n\}$ may follow $\tau^*$ and may have $\tau^*$ followed.

$(C_1',C_2')\in R$, $\mathcal{E}_1,C_1\models \varphi$, $\mathcal{E}_1,C_1'\models_{\eta_1[z_1\mapsto e_1,\cdots,z_n\mapsto e_n]}\psi$, $\mathcal{E}_2,C_2'\models_{\eta_2[z_1\mapsto g_1,\cdots,z_n\mapsto g_n]}\psi$, by Lemma \ref{frag}.3, $\mathcal{E}_2,C_2\models_{\eta_2} \varphi$, since $\varphi$ is closed, $\mathcal{E}_2,C_2\models \varphi$.

($\Leftarrow$)We prove that the relation

\begin{center}
    $$R=\{(C_1,C_2)|\forall\varphi\in\mathcal{L}^{\tau}_p (\mathcal{E}_1,C_1\models\varphi\textrm{ iff }\mathcal{E}_2,C_2\models\varphi)\}$$
\end{center}

is a weak pomset bisimulation.

We proceed by contradiction. For $X\subseteq\hat{E_1}$ is a pomset and $Y\subseteq\hat{E_2}$ is a pomset, let $(C_1,C_2)\in R$, $C_1\xRightarrow{X}C_1'$, $C_2\xRightarrow{Y}C_2'$ and $X\sim Y$ as pomsets, then $(C_1',C_2'\notin R)$. Hence, there exists a closed formula $\psi\in\mathcal{L}^{\tau}_p$, such that $\mathcal{E}_1,C_1'\models\psi$ and $\mathcal{E}_2,C_2'\nvDash\psi$.

Since all PESs are image finite, we assume some step $Y^i$ and some formula $\psi^i$, $\mathcal{E}_1,C_1'\models\psi^i$ and $\mathcal{E}_2,C_2'\nvDash\psi^i$.

Then we can construct a formula

\begin{center}
    $$\varphi=\langle\langle |p_Z|\rangle\rangle(\psi^1\wedge\cdots\wedge\psi^k)$$,
\end{center}

such that $\mathcal{E}_1,C_1\models\varphi$ and $\mathcal{E}_2,C_2\nvDash\varphi$.
\end{proof}

\subsection{HP Logic $\mathcal{L}^{\tau}_{hp}$ and Weak HP-Bisimilarity $\approx_{hp}$}

The syntax of hp logic $\mathcal{L}^{\tau}_{hp}$ is as follows:

\begin{center}
$$\varphi::=\langle\langle |\textbf{x},\overline{\textbf{y}}\ll az|\rangle\rangle\varphi | \varphi\wedge\varphi | \neg\varphi | \textsf{T}$$
\end{center}

\textbf{Lemma \ref{frag}.5.} Let $\mathcal{E}_1$ and $\mathcal{E}_2$ be PESs and let $(C_1,f,C_2)\in\mathcal{C}(\mathcal{E}_1)\overline{\times}\mathcal{C}(\mathcal{E}_2)$, $C_1\in\mathcal{C}(\mathcal{E}_1)$ and $C_2\in\mathcal{C}(\mathcal{E}_2)$, are configurations, and $f:\hat{C_1}\rightarrow\hat{C_2}$ is an isomorphism of pomsets. Then, $R$ is a weak hp-bisimulation and $(C_1,f,C_2)\in R$, iff, for any $\varphi\in\mathcal{L}^{\tau}_{hp}$, $\eta_1\in Env_{\mathcal{E}_1}$ and $\eta_1(fv(\varphi))\subseteq C_1$, $\mathcal{E}_1,C_1\models_{\eta_1}\varphi\Leftrightarrow\mathcal{E}_2,C_2\models_{f\circ\eta_1}\varphi$.

\begin{proof}
($\Rightarrow$)For $(C_1,f,C_2)\in R$ and $R$ is a weak hp-bisimulation, $\eta_1\in Evn_{\mathcal{E}_1}$ such that $\eta_1(fv(\varphi))\subseteq C_1$, to get $\mathcal{E}_1,C_1\models_{\eta_1}\varphi\Leftrightarrow\mathcal{E}_2,C_2\models_{f\circ\eta_1}\varphi$ for all $\varphi\in\mathcal{L}^{\tau}_{hp}$, we need to induct on the structure of $\varphi$.

For the nontrivial case $\varphi=(\langle\langle |\textbf{x},\overline{\textbf{y}}\ll a z|\rangle\rangle\psi$, if $\mathcal{E}_1,C_1\models_{\eta_1}\varphi$, there is an event $e\in\hat{E}_1$, such that $C_1\xRightarrow{e}C_1'$, with $\lambda_1(e)=a$, $\eta_1(\textbf{x})\ll e$, $\eta_1(\textbf{y})\parallel \tau^*e\tau^*$, and $\mathcal{E}_1,C_1'\models_{\eta_1'}\psi$, where $\eta_1'=\eta_1[z\mapsto e]$.

Since $(C_1,f,C_2)\in R$, there exists an event $g\in\hat{E_2}$, such that $C_2\xRightarrow{g}C_2'$, and $(C_1',f',C_2')\in R$ with $f'=f[e\mapsto g]$, we get $\lambda_2(g)=a$, $f(\eta_1(\textbf{x}))\ll g$, and $f(\eta_1(\textbf{y}))\parallel \tau^*g\tau^*$. Since $\eta_1'(fv(\psi))\subseteq\eta_1'(fv(\varphi)\cup\{z\})=\eta_1(fv(\varphi))\cup\{e\}\subseteq \hat{C_1}\cup\{e\}=\hat{C_1'}$, then we get $\mathcal{E}_2,C_2'\models_{f'\circ\eta_1'}\psi$. So, $\mathcal{E}_2,C_2\models_{f\circ\eta_1}\varphi$.

The converse also can be proven analogously.

($\Leftarrow$)We fix $\eta_1:Var\rightarrow \hat{E_1}$. For $e\in\hat{E_1}$, we let $\eta_1(x_e)=e$. For $\hat{C_1}=\{e_1,\cdots,e_n\}$, we let $X_{\hat{C_1}}=\{x_{e_1},\cdots,x_{e_n}\}$.

We prove that the relation

\begin{center}
    $$R=\{(C_1,f,C_2)|\forall\varphi\in\mathcal{L}^{\tau}_{hp}.fv(\varphi)\subseteq X_{\hat{C_1}} (\mathcal{E}_1,C_1\models_{\eta_1}\varphi\textrm{ iff }\mathcal{E}_2,C_2\models_{f\circ\eta_1}\varphi)\}$$
\end{center}

is a weak hp-bisimulation.

We proceed by contradiction. For $e\in\hat{E_1}$, $g\in\hat{E_2}$, let $(C_1,f,C_2)\in R$, $C_1\xRightarrow{e}C_1'$, $C_2\xRightarrow{g}C_2'$ and $\hat{C_1'}\sim \hat{C_2'}$ as pomsets, then $(C_1',f[e\mapsto g],C_2'\notin R)$. Hence, there exists a formula $\psi\in\mathcal{L}^{\tau}_{hp}$ with $fv(\psi\subseteq X_{\hat{C_1'}})$, such that $\mathcal{E}_1,C_1'\models_{\eta_1}\psi$ and $\mathcal{E}_2,C_2'\nvDash_{f'\circ\eta_1}\psi$.

Since all PESs are image finite, we assume some step $Y^i$, some formula $\psi^i$, some $g^i$, and some $f^i=f[e\mapsto g^i]:\hat{C_1'}\rightarrow \hat{C^i_2}$ is an isomorphism of pomsets, $\mathcal{E}_1,C_1'\models_{\eta_1}\psi^i$ and $\mathcal{E}_2,C^i_2\nvDash_{f^i\circ\eta_1}\psi^i$ where $fv(\psi^i)\subseteq X_{\hat{C_1'}}=X_{\hat{C_1}}\cup\{x_e\}$.

Then we can construct a formula

\begin{center}
    $$\varphi=\langle\langle |\textbf{x},\overline{\textbf{y}}\ll a x_e|\rangle\rangle(\psi^1\wedge\cdots\wedge\psi^k)$$,
\end{center}

such that $\mathcal{E}_1,C_1\models_{\eta_1}\varphi$ and $\mathcal{E}_2,C_2\nvDash_{f\circ\eta_1}\varphi$.
\end{proof}

\textbf{Theorem \ref{frag}.6 (Weak Hp-Bisimilarity, Logically).} Let $\mathcal{E}_1$ and $\mathcal{E}_2$ be PESs. Then, $\mathcal{E}_1\approx_{hp}\mathcal{E}_2$ iff $\mathcal{E}_1\equiv_{\mathcal{L}^{\tau}_{hp}}$.

\begin{proof}
($\Rightarrow$)Let $\mathcal{E}_1\approx_{hp}\mathcal{E}_2$, then there is a weak hp-bisimulation $R$ such that $(\emptyset,\emptyset,\emptyset)\in R$. For all $\varphi\in\mathcal{L}^{\tau}_{hp}$, if $\varphi$ is closed, we get $\mathcal{E}_1,\emptyset\models_{\eta_1}\varphi$ iff $\mathcal{E}_2,\emptyset\models_{f\circ\eta_1}\varphi$ for any $\eta_1\in Env_{\mathcal{E}_1}$, that is, $\mathcal{E}_1\models\varphi$ iff $\mathcal{E}_1\models\varphi$, that is, $\mathcal{E}_1\equiv_{\mathcal{L}^{\tau}_{hp}}\mathcal{E}_2$.

($\Leftarrow$) Let $\mathcal{E}_1\equiv_{\mathcal{L}^{\tau}_{hp}}\mathcal{E}_2$. Then, for any closed formula $\varphi\in\mathcal{L}^{\tau}_{hp}$, we get $\mathcal{E}_1,\emptyset\models_{\eta_1}\varphi$ iff $\mathcal{E}_2,\emptyset\models_{\eta_2}\varphi$, then, we get $\mathcal{E}_1,\emptyset\models_{\eta_1}\varphi$ iff $\mathcal{E}_2,\emptyset\models_{\empty\circ\eta_1}\varphi$. So, it says that there exists a weak hp-bisimulation $R$ such that $(\emptyset,\emptyset,\emptyset)\in R$, that is, $\mathcal{E}_1\approx_{hp}\mathcal{E}_2$.

\end{proof}

\section{The Logic $\mathcal{L}^{\tau}$ with Recursion}{\label{recur}}

To express infinite computation, we extend $\mathcal{L}^{\tau}$ with recursion by a fixpoint operator, which is called $\mu\mathcal{L}^{\tau}$.

$\mu\mathcal{L}^{\tau}$ also induces weak hhp-bisimilarity. The solution is similar to that of $\mu\mathcal{L}$ inducing hhp-bisimilarity (which is similar to $\mu$-calculus inducing (interleaving) bisimilarity \cite{MUC}). In the following, we just give the skeleton.

Let $\mathcal{X}^a$ be a set of abstract proposition ranged by $X,Y,\cdots$. A abstract proposition $X$ can be turned into a formula by specifying a name for its free variables as $X(\textbf{x})$, and $|\textbf{x}|=ar(X)$, where $ar(X)$ is the arity of $X$, and $X\in\mathcal{X}$.

\textbf{Definition \ref{recur}.1 (Syntax).} Let $Var$ be a denumerable set of event variables and let $\mathcal{X}$ be a set of propositions. The syntax of $\mu\mathcal{L}^{\tau}$ over labels $\Lambda$ is defined as follows:

\begin{center}
    $$\varphi::=X(\textbf{x})|\textsf{T}|\varphi\wedge\varphi|\neg\varphi|(\textbf{x},\overline{\textbf{y}}\ll az)\varphi|\langle\langle z\rangle\rangle\varphi|\mu X(\textbf{x}).\varphi$$,
\end{center}

where for formula $\mu X(\textbf{x}).\varphi$, $X$ must occur positively in $\varphi$ to ensure the existence of the fixpoint, $fv(\varphi)=\textbf{x}$.

The free variables of a formula $\varphi$ in $\mu\mathcal{L}^{\tau}$ are added the following two clauses: $fv(X(\textbf{x}))=\textbf{x}$ and $fv(\mu X(\textbf{x}).\varphi)=\textbf{x}$.

The greatest fixpoint operator can be defined as $\nu X(\textbf{x}).\varphi=\neg(\mu X(\textbf{x}.\neg\tilde{\varphi}))$, where $\tilde{\varphi}$ is the formula obtained replacing any occurrence of $X$ in $\varphi$ with $\neg X$.

The set of free propositions in a formula $\varphi$ in $\mu\mathcal{L}$ as $fp(\varphi)$, is defined inductively, we just enumerate the non-trivial ones in $\mu\mathcal{L}^{\tau}$

$fp((\textbf{x},\overline{\textbf{y}}\ll az)\varphi)=fp(\langle\langle z\rangle\rangle\varphi)=fp(\varphi)$

Let $\mathcal{E}$ be a PES. A proposition environment is a function $\pi:\mathcal{X}\rightarrow 2^{\mathcal{C}(\mathcal{E})\times Env_{\mathcal{E}}}$, same as that in $\mu\mathcal{L}$. And the conclusions about the proposition environments $\pi$ in $\mu\mathcal{L}$ also hold in $\mu\mathcal{L}^{\tau}$.

Let $\mathcal{E}$ be a PES. The denotation of the non-trivial formula in $\mu\mathcal{L}^{\tau}$ defined inductively as follows:

\begin{enumerate}
  \item $\{|(\textbf{x},\overline{\textbf{y}}\ll az)\varphi|\}^{\mathcal{E}}_{\pi}=\{(C,\eta)|(C,\eta)\in lp((\textbf{x},\overline{\textbf{y}}\ll az)\varphi)\textrm{ and } \exists e\in\hat{E(C)}\textrm{ such that } e\frown \eta(fv(\varphi)\backslash\{z\})\wedge \lambda(e)=a \wedge \eta(\textbf{x})\ll e \wedge \eta(\textbf{y})\parallel(\tau^* e \tau^*)\wedge (C,\eta[z\mapsto e])\in \{|\varphi|\}^{\mathcal{E}}_{\pi}\}$
  \item $\{|\langle\langle z\rangle\rangle\varphi|\}^{\mathcal{E}}_{\pi}=\{(C,\eta)|C\xRightarrow{\eta(z)} C' \wedge (C',\eta)\in \{|\varphi|\}^{\mathcal{E}}_{\pi}\}$
  \item $\{|\mu X(\textbf{x}).\varphi|\}^{\mathcal{E}_{\pi}}=lfp(f)$
\end{enumerate}

Where $lfp(f)$ is the least fixed point of the function $f:2^{lp(X(\textbf{x}))}\rightarrow 2^{lp(X(\textbf{x}))}$. Note that, the least fixed point of $f$ still exists by Knaster-Tarski theorem, Since $\pi[X(\textbf{x})\mapsto S]$ for $S\subseteq lp(X(\textbf{x}))$ only considers event $e\in\hat{E}$.

Let $\mu\mathcal{L}^{\tau,\infty}$ be the extension of $\mu\mathcal{L}^{\tau}$ with infinite conjunctions $\wedge$.

Then the lemma about fixpoint unfolding via approximants in $\mu\mathcal{L}^{\tau,\infty}$ still holds. Its says that: Let $\mathcal{E}$ be a PES. for any formula $\mu X(\textbf{x}).\varphi$ in $\mu\mathcal{L}^{\tau,\infty}$, there exists an ordinal $\alpha$ such that $\{|\mu X(\textbf{x})\varphi|\}^{\mathcal{E}}_{\pi}=\{|\mu^{\alpha}X(\textbf{x}).\varphi|\}^{\mathcal{E}}_{\pi}$.

\textbf{Theorem \ref{recur}.2 (Invariance of Logical Equivalence).} The logical equivalences of $\mathcal{L}^{\tau}$ and $\mu\mathcal{L}^{\tau}$ coincide with $\approx_{hhp}$.

\begin{proof}
$\equiv_{\mu\mathcal{L}^{\tau}}$ implies $\equiv_{\mathcal{L}^{\tau}}$, $\equiv_{\mathcal{L}^{\tau}}$ implies $\approx_{hhp}$, so, $\equiv_{\mu\mathcal{L}^{\tau}}$ implies $\approx_{hhp}$.

$\approx_{hhp}$ implies $\equiv_{\mathcal{L}^{\tau,\infty}}$, the above lemma says that for any closed formula in $\mu\mathcal{L}^{\tau,\infty}$, there exists an equivalent formula in $\mathcal{L}^{\tau,\infty}$, so, $\equiv_{\mathcal{L}^{\tau,\infty}}$ implies $\equiv_{\mu\mathcal{L}^{\tau}}$. So, $\approx_{hhp}$ implies $\equiv_{\mu\mathcal{L}^{\tau}}$.
\end{proof}

The logical equivalences of fragments of $\mu\mathcal{L}^{\tau}$ also coincide with corresponding weakly concurrent bisimilarity. That is, $\equiv_{\mathcal{L}^{\tau}_{HM}}$ and $\equiv_{\mu\mathcal{L}^{\tau}_{HM}}$ coincide $\approx_{HM}$, $\equiv_{\mathcal{L}^{\tau}_{s}}$ and $\equiv_{\mu\mathcal{L}^{\tau}_{s}}$ coincide $\approx_{s}$, $\equiv_{\mathcal{L}^{\tau}_{p}}$ and $\equiv_{\mu\mathcal{L}^{\tau}_{p}}$ coincide $\approx_{p}$, and also $\equiv_{\mathcal{L}^{\tau}_{hp}}$ and $\equiv_{\mu\mathcal{L}^{\tau}_{hp}}$ coincide $\approx_{hp}$.

\section{Conclusions}{\label{conc}}

We firstly find the existence of silent event $\tau$ in true concurrency (named weakly true concurrency) by defining weakly true concurrent behaviors and weakly true concurrent logics. Based on Paolo Baldan and Silvia Crafa's comprehensive work \cite{LTC1} \cite{LTC2} on (strongly) true concurrency, we find the correspondence between weakly true concurrent equivalences and weakly true concurrent logics.

\newpage

%

\label{lastpage}

\end{document}